\documentclass[5p,times,procedia]{elsarticle}
\usepackage{ecrc}
\usepackage{amsmath}

\volume{00}

\firstpage{1}

\journalname{Procedia CIRP}

\runauth{Rüppel et al.}

\jid{trpro}

\usepackage{amssymb}

\usepackage[figuresright]{rotating}

\usepackage{color,soul}

\usepackage{siunitx}

\usepackage{makecell} 

\usepackage[bookmarks=false]{hyperref}
    \hypersetup{colorlinks,
      linkcolor=blue,
      citecolor=blue,
      urlcolor=blue}

\usepackage{etoolbox}

\makeatletter
\pretocmd{\NAT@citexnum}{\@ifnum{\NAT@ctype>\z@}{\let\NAT@hyper@\relax}{}}{}{}
\makeatother

\begin{document}
\begin{frontmatter}



\dochead{55th CIRP Conference on Manufacturing Systems}%

\title{Mass Loss and Displacement Modeling for Multi-Axis Milling}


\author[a]{Adrian Karl R\"uppel\corref{cor1}}

\author[b]{Patrick Ochudlo}
\author[a]{Mathias Bickel}
\author[b]{Sebastian Stemmler}
\author[a,c]{Thomas Bergs}
\author[b]{Dirk Abel}

\address[a]{Laboratory for Machine Tools and Production Engineering (WZL), RWTH Aachen University, Campus-Boulevard 30, 52074 Aachen, Germany}
\address[b]{Institute of Automatic Control (IRT), RWTH Aachen University, Campus-Boulevard 30, 52074 Aachen, Germany}
\address[c]{Fraunhofer Institute for Production Technology (IPT), Steinbachstraße 17, 52074 Aachen, Germany}

\aucores{* Corresponding author. Tel.: +49-241-80-28020 ; fax: +49-241-80-22293. {\it E-mail address:} a.rueppel@wzl.rwth-aachen.de}

\begin{abstract}
During the cutting process, material of the workpiece is continuously being removed by the cutting tool, which results in a reduction of mass as well as a displacement in the center of the workpiece mass. When using workpiece sided force sensors, such as table dynamometers, the total mass and the displacement of the center of mass affects the force measurement due to gravitational and inertial effects. The high flexibility of the milling process leads to a complex change of volume and mass and necessitates the consideration of the engagement conditions between tool and workpiece along the tool path in order to estimate changes in mass and center of mass. This paper proposes a method for estimating the mass loss and the displacement of the center of mass during multi-axis milling processes. In this method the tool gets numerically sliced along the tool axis and the workpiece removal for each slice along an arbitrary tool path gets calculated. To validate the mass loss model, experiments in both three-axis milling as well as multi-axis milling processes have been conducted. Since it is difficult to measure the center of mass, validation for the displacement of the center of mass was done by comparison with data extracted from CAD. The results show good agreement between the simulated and measured mass loss using the proposed approach. 
\end{abstract}

\begin{keyword}
milling; mass change; center of mass




\end{keyword}

\end{frontmatter}



\section{Introduction and state of the art}
\label{sec:Introduction}
In machining, process forces determine productivity, process safety and product quality. An accurate measurement of dynamic process forces enables force monitoring, control and optimizing of cutting processes \cite{Ulsoy.1993}. Measuring process forces is both costly and often difficult to establish. The combination of high loads, high temperatures, and high process dynamics requires special force measurement devices. Piezo-electric force dynamometers are of the most widely used measuring devices today \cite{Teti.2022}. They can be mounted both tool and workpiece sided. While tool sided dynamometers are more flexible concerning multi-axis milling, their usage is limited in terms of tool size. Furthermore, the measuring point is at the clamping of the tool, far away from the tool-workpiece engagement, requiring the consideration of tool bending to determine dynamic process forces.

Workpiece-sided devices, such as table dynamometers, allow the measurement of process forces closer to the engagement point. However, table dynamometers are currently limited for three axis milling processes, as a movement of the table also moves the dynamometer and the workpiece and resulting in gravitational and inertial influences on the measured forces. To isolate the process forces from the other influences, a correction of the raw force signal is necessary. Correction of measured forces in multi-axis milling requires consideration of gravitational and inertial components. The mass of the workpiece, which is constantly changing during the cutting process, affects both force components. Additionally, the center of mass, which is also constantly changing, alters the inertial components as well. The intensity of the effect depends on the total change in mass, the angle between the vector of the gravitational force and the dynamometer zero orientation, and the acceleration of the machine table. Especially high total material removals lead to significant disturbances in the measured forces. To determine all force components and correct the measured forces, precise information about both mass and center of mass of the workpiece is necessary for every position along the tool path during the milling process.

Correction of process forces in literature is mostly conducted by determining the transfer behavior of the dynamometer without considering the workpiece \cite{RicardoCastro.2006, Scippa.2015, Wan.2016, Totis.2020}. \citeauthor{Klocke.2008} \cite{Klocke.2008} studied the influences of the workpiece mass on the transfer behavior of piezo-electric dynamometers. The authors showed that the mass of the workpiece as well as the clamping situation of the dynamometer influenced the transfer function. Especially high modes are significantly affected by different workpiece masses. However, only three-axis milling was investigated and mass loss was not modeled.

A correction of gravitational and inertial effects in a multi-axis milling-like scenario was conducted by \citeauthor{Rekers.2019} \cite{Rekers.2019} by mounting a dynamometer on top of a serial robot. The necessity of a mass change model is mentioned, however, not established due to the experimental design without material removal. \citeauthor{Klocke.2014} \cite{Klocke.2014} mentioned the possibility to model mass loss by means of the material removal rate, but did not establish a model as well reasoning mass loss is not significant in their researched finishing processes. While the suggested modeling approach based on material removal rate might give information about the change in mass during machining, it is not sufficient for determining displacements of the center of mass. 

Changes in mass also affect the dynamic behavior of the workpiece. To determine structural modes during the milling process, \citeauthor{Wang.2020} \cite{Wang.2020} extracts information on mass change for tool path segments from CAD. This methodology provides information on mass loss, however, is not possible for every position along the tool path with reasonable effort.

To conclude the state of the art, no method to determine both mass loss and displacement of the center of mass in machining has been established according to the current state of knowledge of the authors. While some works consider the workpiece mass in the transfer behavior of dynamometer based force measurement changes in mass during machining are not considered. Explicit information about mass and center of mass, however, are necessary to compensate gravitational and inertial effects during multi-axis milling using table sided force sensors. The quantity of the effects depend on the relative amount of machined material and the displacement of the center of mass during machining.

This paper introduces a method to determine the mass and the center of mass of the workpiece for every position along an arbitrary tool path in multi-axis milling. To extract this information from CAD requires the design of 3D models for all machining states along the tool path for every tool position, which is very time consuming and, therefore, not applicable. The aim of the work was the enhancement of a model predictive force control to multi-axis milling processes, which is related on a piezo-electric table dynamometer \cite{Schwenzer.2022b}. The control system requires accurate force measurement with a high bandwidth to identify a force model online.

\section{Approach}
\label{sec:Approach}

Mass loss and displacement of the center of mass both can be determined a-priori by approximating the volume of the machined material, assuming homogeneous material density. However, for determining the displacement of center of mass, information on the total value of the volume as well as the distribution in space of the removed material are necessary. 

The flexibility of milling processes, especially during multi-axis milling, leads to complex volumetric changes during machining. Throughout this work multi-axis milling is defined as milling with tilted workpiece resulting in the tool rotation axis being not perpendicular to the workpiece normal plane and to non-constant engagement conditions between workpiece and tool.

A material removal simulation is conducted to determine precise information about the tool-workpiece-engagement. The simulation is conducted using the software dPart developed by Fraunhofer Institute for Production Technology (IPT), which offers a dexel-based numerical engagement simulation and provides entry angles $\varphi_{in}$ and exit angles $\varphi_{ex}$ of the cutting edges along arbitrary tool paths \cite{Ganser.2021}. To account for helical milling tools, the tool gets further numerically sliced into disk elements along the tool rotation axis \cite{Sutherland.1986}. 

The disk height determines the accuracy of the model, but also increases the computation time. According to \citeauthor{Schwenzer.2022} \cite{Schwenzer.2022}, the disk height should be chosen to result in an angular resolution of the exact position of the cutting edges less than $\Delta\varphi_{max} = \SI{0.5}{\degree}$. Therefore, the disk height was chosen to be \SI{0.1}{\milli\meter}, which results in an angular resolution of the cutting edge of $\Delta\varphi=\SI{0.4}{\degree}$.

\begin{figure}[!h]
    \centering
	\graphicspath{{Figures/}}
    \includegraphics[width=\columnwidth]{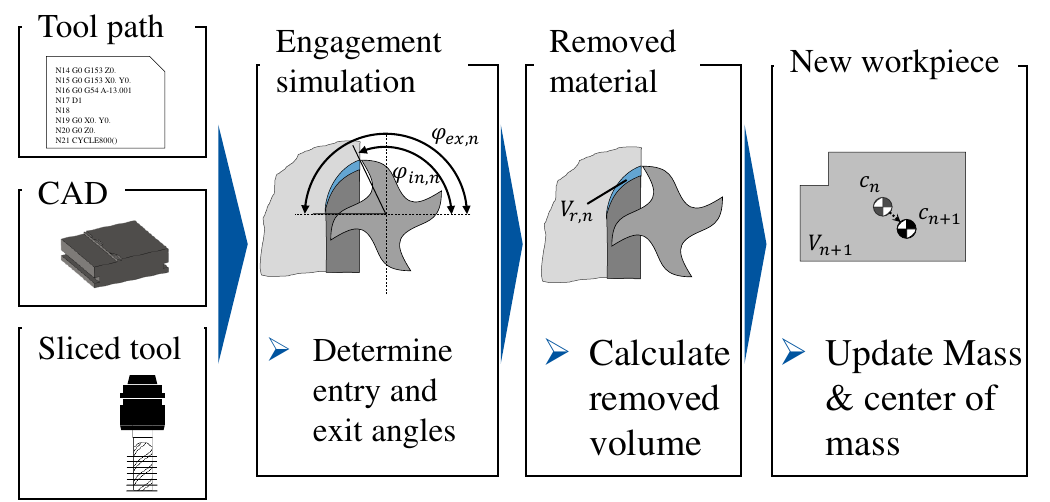}
    \caption{General methodology followed by in the paper. The current position of the tool along the full tool path is denoted with an index $n$.}
    \label{fig:Methodology}
\end{figure}

The total removed volume $V_r$ as well as the total volume of the workpiece $V_n$ and the spacial position of the center of mass $c_n$ can be calculated by summing up all slices for every position along the tool path. The mass and center of mass can be further calculated by considering the material density. The tool moves along the tool path from one position $n$ to the next position $n+1$. Figure~\ref{fig:Methodology} shows the full methodology. The mass change model as well as the experimental setup are further described in the next subsections.

\subsection{Mass change model}
\label{sec:Mass}

During machining, the mass $m_{WP}$ and position $\mathbf{c}_{WP}$ of the center of mass in space of the workpiece $WP$ is constantly changing along the tool path due to material being removed by the tool. The workpiece mass $m_{WP,n+1}$ at the next tool position $n+1$ results from the current mass $m_{WP,n}$ at tool position $n$ reduced by the removed mass $m_{r,n}$ of the current cut and was calculated by

\begin{equation}
	m_{WP,n+1} = m_{WP,n} - m_{r,n}.
\end{equation}

The mass of the removed material was calculated by considering the density $\rho$ of the workpiece material and the removed volume $V_{r,n}$ as follows

\begin{equation}
	m_{r,n} = \rho V_{r,n}.
\end{equation}

The center of mass $\mathbf{c}_{WP,n+1}$ at the next position is defined as

\begin{equation}
	\begin{pmatrix}
		x_{n+1} & y_{n+1} & z_{n+1}
	\end{pmatrix}^T	 
\end{equation}

and results according to


\begin{equation}
	\mathbf{c}_{WP,n+1} = \frac{1}{m_{WP,n+1}} (\mathbf{c}_{WP,n} m_{WP,n} - \mathbf{c}_{r,n} m_{r,n})
\end{equation}


with the position $\mathbf{c}_{WP,n}$ of the center of mass at the current position and the position $\mathbf{c}_{r,n}$ of the center of mass of the removed material. By assuming homogeneous material, the center of mass equals the center of volume $\mathbf{v}_{WP}$ and the new center of mass can be calculated as

\begin{equation}
	\mathbf{c}_{WP,n+1} = \mathbf{v}_{WP,n+1} = \frac{1}{V_{WP,n+1}} (\mathbf{c}_{WP,n} V_{WP,n} - \mathbf{c}_{r,n} V_{r,n}).
\end{equation}

\begin{figure}[!h]
    \centering
    \def\svgwidth{\columnwidth}
	\graphicspath{{Figures/}}
    \input{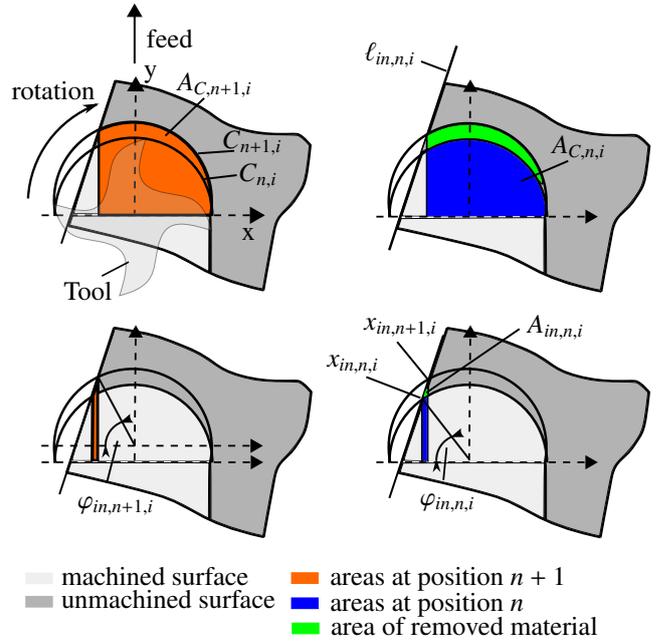}
    \caption{Different areas of the removed material at the same tool position for one slice. The area $A_{ex,n,i}$ is defined accordingly to $A_{in,n,i}$ at the exit of the tooth but not relevant in the shown engagement situation and, therefore, not displayed.}
    \label{fig:removedMaterial}
\end{figure}

For approximation of the volume $V_{r,n}$ and center of mass $\mathbf{c}_{r,n}$ of the removed material, the volume and center of mass of the slices $i$ of the tool are summed up. The problem reduces to the calculation of the area $A_{r,n,i}$ of the removed material of every slice, according to

\begin{equation}
	V_{r,n,i} = A_{r,n,i} b_i 
\end{equation}

with the slice height $b_i$.  The resulting center of mass is calculated accordingly. For the area of the removed material of each slice, the information of the engagement between tool and workpiece is being used. The assumption of the cutting edges moving in circular paths is taken, which is a well established simplification \cite{Kronenberg.1969}. Figure~\ref{fig:removedMaterial} shows the calculated areas of each slice assuming circular cutting edge movement.

The area under the tooth path at the next position is denoted by $A_{C,n+1,i}$, the area under the tooth path at the current position by $A_{C,n,i}$. The areas $A_{in,n,i}$ and $A_{ex,n,i}$ mark the interpolations between the two positions at entry and exit of the tooth and are dependent on the engagement conditions. The entry angle $\varphi_{in,n,i}$ of the current cut, the entry angle $\varphi_{in,n+1,i}$ of the next cut, the exit angle $\varphi_{ex,n,i}$ of the current cut, and the exit angle $\varphi_{ex,n+1,i}$ of the next cut determine the overlap of the tool and the workpiece. The removed area $A_{r,n,i}$ of the slice can be calculated according to 

\begin{equation}
	A_{r,n,i} = A_{C,n+1,i} - A_{C,n,i} \pm A_{in,n,i} \pm A_{ex,n,i}.
\end{equation}

The four areas can be calculated as follows

\begin{align}
	A_{C,n+1,i} &= \int_{x_{in,n+1,i}}^{x_{ex,n+1,i}} C_{n+1,i}(x) dx \\
	A_{C,n,i} &= \int_{x_{in,n+1,i}}^{x_{ex,n+1,i}} C_{n,i}(x) dx \\
	A_{in,n,i} &= \Bigg|\int_{x_{in,n,i}}^{x_{in,n+1,i}} \ell_{in,n,i}(x) dx\Bigg| - \Bigg|\int_{x_{in,n,i}}^{x_{in,n+1,i}} C_{n,i}(x) dx\Bigg| \\
	A_{ex,n,i} &= \Bigg|\int_{x_{ex,n,i}}^{x_{ex,n+1,i}} \ell_{ex,n,i}(x) dx\Bigg| - \Bigg|\int_{x_{ex,n,i}}^{x_{ex,n+1,i}} C_{n,i}(x) dx \Bigg|
\end{align}

The areas are determined by the circular path $C_{n,i}$ of the cutting edge of the current cut, the circular path $C_{n+1,i}$ of the cutting edge of the next cut, the interpolation line $\ell_{in,n,i}$ between the tooth entry points, and the interpolation line $\ell_{ex,n,i}$ between the tooth exit points. The integration borders are the intersection point ${x_{in,n,i}}$ between $\ell_{in,n,i}$ and $C_{n,i}$, the intersection point ${x_{in,n+1,i}}$ between $\ell_{in,n,i}$ and $C_{n+1,i}$ , the intersection point ${x_{ex,n,i}}$ between $\ell_{ex,n,i}$ and $C_{n,i}$, and the intersection point ${x_{ex,n+1,i}}$ between $\ell_{ex,n,i}$ and $C_{n+1,i}$. In the studied case down milling is assumed, however, the methodology is also applicable for up milling. 

The model requires an initial value for volume $V_{WP,1}$ and center of mass $\mathbf{c}_{WP,1}$ of the raw workpiece, which can be extracted from the CAD file. For later usage in the control system, the information on mass and center of mass of every position along the tool paths are calculated a-priori and saved in a look up table.

\subsection{Experimental Setup}
\label{sec:ExperimentalSetup}

\begin{figure}[!h]
    \centering
    \def\svgwidth{\columnwidth}
	\graphicspath{{Figures/}}
    \input{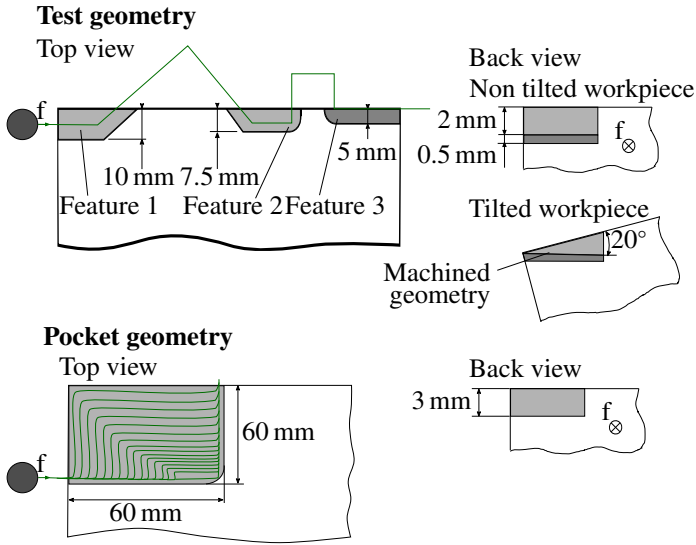}
    \caption{Machined geometries}
    \label{fig:machinedGeometry}
\end{figure}

To validate the developed model under various engagement conditions, a geometry introduced by \citeauthor{Schwenzer.2022b} \cite{Schwenzer.2022b} is used. It covers abrupt and continuous changes in entry and exit angles as well as different width and depth of cuts. Figure~\ref{fig:machinedGeometry} shows the machined geometry. For experimental validation on both three-axis and multi-axis milling scenarios, the workpiece gets machined while the workpiece is flat, i.e. perpendicular to the tool rotation axis, as well as with a tilted workpiece.

Furthermore, to validate the feasibility of the model in more realistic scenarios, a pocket milling operation was conducted additionally. The experiments are conducted on a Mazak VARIAXIS i-600 5-axis machining center using a two fluted end mill of the type {JS553030G2R050.0Z3-SIRA} from Seco Tools GmbH and a workpiece of the material AlZnMgCu1.5 with a density of \SI{2.81}{\gram\per\cubic\centi\meter} \cite{HansErichGemmel&Co.GmbH.2011}. No cutting fluid was used. The masses were measured using a scale of type Precision Balance {XPE3003S} from Mettler Toledo.

\section{Results and discussion}
\label{sec:Results}

To validate the mass loss model experiments as well as comparison with a CAD model using SOLIDWORKS where conducted. The validation of displacements in the center of mass was performed by comparison with CAD data of the displacements of the centers only. For the test geometry shown in Figure~\ref{fig:machinedGeometry} simulative validation was conducted for all three geometry features separately. The resolution along the tool path was chosen to be \SI{0.5}{\milli\meter} based on previous experiments. Table~\ref{tab:ResultsThreeAxisMass} shows the masses losses $\Delta m_{mod}$ calculated by the mass loss model, the manually calculated masses losses $\Delta m_{CAD}$ extracted from CAD, the measured mass losses $\Delta m_{meas}$, as well as the relative error $e_{CAD}$ of the model compared to CAD and the relative error $e_{meas}$ compared to measurement. The measurements where only taken for the full geometry and not for single features.

\begin{table}[!h]
	\centering
	\renewcommand\cellalign{lc}
	\begin{tabular}{l c c c c c}
		Geometry & $\Delta m_{mod}$ & $\Delta m_{CAD}$ & $\Delta m_{meas}$ & $e_{CAD}$ & $e_{meas}$ \\
		 & [g] &   [g] &  [g] & [\%] & [\%]	\\
		\Xhline{3\arrayrulewidth}		
		\makecell{Test geometry \\ feature 1} & 1.65 & 1.58  & - & 4.7 & - \\
		\hline
		\makecell{Test geometry \\ feature 2} & 1.24 & 1.23  & - & 0.9 & - \\
		\hline
		\makecell{Test geometry \\ feature 3} & 1.11 & 1.12  & - & 0.9 & - \\
		\hline
		\makecell{Test geometry \\ total} & 4.01 & 3.93  & 4.18 & 1.9 & 4.2 \\
		\Xhline{2\arrayrulewidth}
		Pocket & 28.4 & 27.13  & 28.51 & 4.7 & 0.4 \\
	\end{tabular}
	\caption{Modeled, CAD extracted, and measured mass loss in three-axis milling}
	\label{tab:ResultsThreeAxisMass}
\end{table}

The model agreed well with the CAD data and the measured masses with relative errors less than $\SI{5}{\percent}$ in three axis milling. Differences in the test geometry between CAD and model in feature~1 result from slightly different exit angle of the tool, which was due to numerical variations in the engagement simulation. For the pocket milling operation relative errors between CAD and mass loss model were slightly higher, however, under $\SI{1}{\percent}$ compared to the measurement.

In multi-axis milling with tilted workpiece similar results are shown in Table~\ref{tab:ResultsMultiAxisMass}. No pocket has been machined with tilted workpiece.

\begin{table}[!h]
	\centering
	\renewcommand\cellalign{lc}
	\begin{tabular}{l c c c c c}
		Geometry & $\Delta m_{mod}$ & $\Delta m_{CAD}$ & $\Delta m_{meas}$ & $e_{CAD}$ & $e_{meas}$ \\
		 & [g] &   [g] &  [g] & [\%] & [\%]	\\
		\Xhline{3\arrayrulewidth}		
		\makecell{Test geometry \\ feature 1} & 1.39 & 1.33  & - & 5.1 & - \\
		\hline
		\makecell{Test geometry \\ feature 2} & 0.77 & 0.8  & - & 4.3 & - \\
		\hline
		\makecell{Test geometry \\ feature 3} & 0.57 & 0.62  & - & 7.8 & - \\
		\hline
		\makecell{Test geometry \\ total} & 2.73 & 2.75  & 2.99 & 0.6 & 8.8 \\
	\end{tabular}
	\caption{Modeled, CAD extracted, and measured mass loss in multi-axis milling}
	\label{tab:ResultsMultiAxisMass}
\end{table}

During multi-axis milling, the accuracy of the model was slightly worse resulting in an maximum error of $\SI{7.8}{\percent}$ compared to the CAD data in feature~3 of the test geometry. However, for the total geometry the errors canceled each other out resulting in only $\SI{0.6}{\percent}$. The error compared to the measurement was slightly higher than in three-axis milling resulting in $\SI{8.8}{\percent}$.

\begin{figure}[!h]
    \centering
    \def\svgwidth{\columnwidth}
	\graphicspath{{Figures/}}
    \input{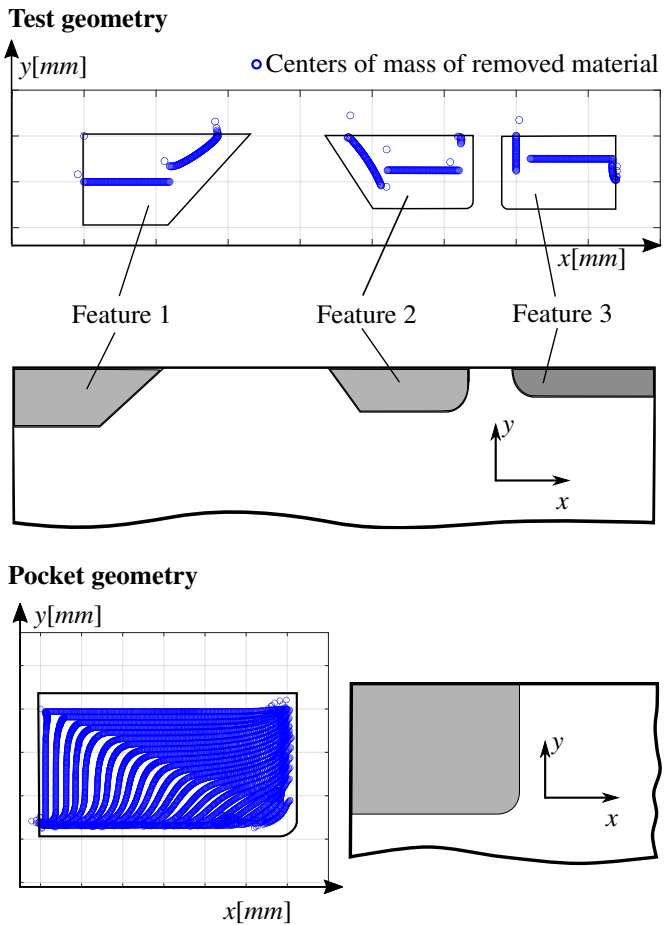}
    \caption{Center of mass of the removed material on every position in three-axis milling (top view)}
    \label{fig:centers_of_gravity}
\end{figure}

The displacements of the center of mass have been validated by comparing the model via CAD data of the separate features as well as the total test geometry and the pocket operation. Initial center of mass for the math. model has been extracted from CAD. Figure~\ref{fig:centers_of_gravity} shows the modeled center of mass of the removed material for every point along the tool path.

The positions of the center of mass agreed with the expected positions in the vast majority of the tool path. However, some calculated centers of the test geometry are outside the machined area. Further validation has been conducted by comparing model and CAD data and evaluating the error $e_{CAD}$. Table~\ref{tab:resultsCenterOfMassThreeAxis} shows the results for the displacement in three-axis milling. $\Delta c$ denotes the euclidean distance between the center of mass before and after machining.

\begin{table}[!h]
	\centering
	\renewcommand\cellalign{lc}
	\begin{tabular}{l c c c}
		Geometry & $\Delta c_{mod}$ & $\Delta c_{CAD}$ & $e_{CAD}$ \\
		 & [mm] &   [mm] & [\%]	\\
		\Xhline{3\arrayrulewidth}		
		\makecell{Test geometry \\ feature 1} & 0.0369 & 0.0356  & 3.6 \\
		\hline
		\makecell{Test geometry \\ feature 2} & 0.0162 & 0.0154 & 4.7 \\
		\hline
		\makecell{Test geometry \\ feature 3} & 0.0248 & 0.0254 & 2.2 \\
		\hline
		\makecell{Test geometry \\ total} & 0.0483 & 0.0476 & 1.4  \\
		\Xhline{2\arrayrulewidth}
		Pocket & 0.6641 & 0.6176 & 4.3  \\
	\end{tabular}
	\caption{Deviations in center of mass calculated by model and by manual extraction from CAD file in three-axis milling.}
	\label{tab:resultsCenterOfMassThreeAxis}
\end{table}

The model showed good agreement with the CAD data. Errors were between $\SI{1.4}{\percent}$ and $\SI{4.7}{\percent}$ for the test geometry and $\SI{4.3}{\percent}$ for the pocket milling operation. 

Accordingly, Table~\ref{tab:resultsCenterOfMassMultiAxis} shows the results for multi-axis milling with the workpiece tilted.

\begin{table}[!h]
	\centering
	\renewcommand\cellalign{lc}
	\begin{tabular}{l c c c}
		Geometry & $\Delta c_{mod}$ & $\Delta c_{CAD}$ & $e_{CAD}$ \\
		 & [mm] &   [mm] & [\%]	\\
		\Xhline{3\arrayrulewidth}		
		\makecell{Test geometry \\ feature 1} & 0.0313 & 0.0299  & 4.4 \\
		\hline
		\makecell{Test geometry \\ feature 2} & 0.0097 & 0.0107  & 9.3 \\
		\hline
		\makecell{Test geometry \\ feature 3} & 0.0128 & 0.0143  & 10.5 \\
		\hline
		\makecell{Test geometry \\ total} & 0.034 & 0.0341 & 0.2  \\
	\end{tabular}
	\caption{Center of mass calculated by model and by manual extraction from CAD file in multi-axis milling}
	\label{tab:resultsCenterOfMassMultiAxis}
\end{table}

Errors for multi-axis milling were again slightly higher than for the three axis operation. A maximum error of $\SI{10.5}{\percent}$ occured for feature 3. However, the total geometry was approximated very well with an error of only $\SI{0.2}{\percent}$, showing again that errors cancel each other out during longer operations.

Overall the mass loss and displacements of the center of mass were predicted with little errors. Most of the results showed very good agreement between mathematical model and CAD data. However, numerical variations coming from the dexel-based engagement simulation lead to smaller errors. Numerical errors also occurred in the calculation of the center of mass of the removed material at a few positions, leading to the center being calculated outside of the machined area. Results showed improvement of the model by increase of machined material suggesting errors canceling out each other during longer operations. A comparatively bigger deviation in mass loss between CAD data and model in feature 1 of the test geometry in three-axis milling resulted from a slightly different exit angle in the engagement simulation. The much longer and more complex operation of the pocket milling operation still showed the feasibility of the model. The approximation of the helical milling tool by single disk elements lead to higher deviations for multi-axis milling. 

The errors between simulations and measurements also resulted due to differences in the micro geometry of the tool edges which is not being accounted for by the model. Furthermore, remaining burrs at the machined workpiece also lead to little errors of the model. Results showed an improvement of the model when more material was machined reducing relative influence of burrs and micro geometry. 

To improve the model even more, spacial resolution along the tool path as well as number of disk elements along the tool rotation axis can be increased. Both result in smaller numerical errors but increase the calculation time significantly. The proposed resolutions have been chosen as a trade-off both accuracy and calculation time.

\section{Conclusion}
\label{sec:Conclusion}
This work introduced a method for modeling mass loss and displacement of the center of mass along arbitrary tool paths in milling. Information about mass and center of mass is required to correct gravitational and inertial effects on force measurements in multi-axis milling using table dynamometers. The overall methodology required an engagement simulation of the tool workpiece engagement. The tool was sliced into disk elements along its rotation axis and entry and exit angles of all slices where calculated for every point along the tool path. This information was fed in a geometrical model to calculate the removed material for every point along the tool paths. Using this information the total mass and the center of mass for every point has been calculated by integration of the teeth cutting paths. 

The model showed little errors compared to CAD data and slightly higher errors compared to measured mass loss. Errors are mostly due to numerical effects as well as the fact, that remaining burrs and micro geometry where not modeled. Further improvement of numerical inaccuracies can be done by modeling micro geometry and increasing the resolution along the tool path as well as decreasing the height of the disk elements along the tool rotation axis. Overall the model is adequate for the desired purpose allowing a-priori information on mass loss and displacement of center of mass.

The developed model is a requirement for force measurement using table dynamometers in multi-axis milling processes, which will be published in a future work. The corrected force measurement is necessary for model predictive force control for multi-axis milling.


\section*{Acknowledgements}

The presented research was funded by the Deutsche Forschungsgemeinschaft (DFG, German Research Foundation)—450794086.



%


\bibliography{myLit}
\bibliographystyle{elsarticle-num-names}



%
%
%
%
%
%
%

\clearpage\onecolumn

\normalMode

%
%
%
%

\end{document}